\documentstyle [preprint,prb,aps,eqsecnum]{revtex}

\begin{document}
\draft
\title{Transport properties of one-dimensional
 interacting fermions in aperiodic potentials}
\author{ Juan Carlos Chaves\cite{email1} and Indubala I.
Satija\cite{email2} }
\address{
 Department of Physics and\\
 Institute for Computational Sciences and Informatics,\\
 George Mason University,\\
 Fairfax, VA 22030}
\date{\today}
\maketitle
\begin{abstract}
Motivated by the existence of metal-insulator transition in
one-dimensional non-interacting fermions in quasiperiodic and
pseudorandom potentials, we studied interacting 
spinless fermion
models using exact many-body Lanczos diagonalization techniques.
Our main focus was to understand the 
effect of the fermion-fermion interaction
on the transport properties of aperiodic systems.
We calculated the ground state energy and 
the Kohn charge stiffness $D_c$.
Our numerical results indicate 
that there exists a region in the 
interaction strength parameter space where the system
may behave differently from the 
metallic and insulating phases.
This intermediate phase 
may be characterized
by a power law scaling of the charge stiffness constant
in contrast to the localized phase
where $D_c$ scales exponentially with the size of the system.
\end{abstract}
\section{Introduction}
Understanding the transport properties of interacting
many-fermion disordered systems
has been one of the most challenging problems in recent years.
Non-perturbative effects are introduced by strong
fermion-fermion correlations occurring in these systems
and the available numerical
tools are plagued by the exponential increase
of the Hilbert space as larger and larger systems are studied.

One-dimensional (1D) interacting fermion systems have emerged
as useful models
since exact solutions exist for some of them. 
Additionally, 
non-interacting 1D models, in $aperiodic$ (such as quasiperiodic
or random) potentials
are known to exhibit interesting phase diagrams.
In case of random potentials,
the non-interacting fermion systems exhibit Anderson
localization\cite{Anderson79}. Anderson insulator is characterized by a gapless point
spectrum. This phase should be distinguished from
the one occurring in a Mott insulator which is characterized
by a finite gap
originated in fermion-fermion interaction effects. The 
quasiperiodic
and deterministic aperiodic models
exhibit metal-insulator 
transition\cite{Sokoloff,Hiramoto92,Satija94,Aperiodic}.
In addition, these aperiodic models exhibit a phase 
which is
intermediate between the metallic and the
insulating phase and has been christened
as {\it critical}. Critical behavior is characterized by
{\it multifractal}
quantum 
states and energy spectrum\cite{Sokoloff,Hiramoto92,Satija94,Aperiodic}.
The transport properties for
this phase, for example, resistance, are known to be 
oscillating\cite{Liu86,Cruz93}. 
Unlike the Anderson localized phase characterized by
the exponential decay of the single particle fermion wave function,
the critical states exhibit at most power law decay.

The possible existence of a phase which is in-between metallic and
insulating
in an {\it aperiodic} many-body fermion systems
is a completely open problem.
In this paper we present some preliminary
results regarding this interesting problem in spinless many-body fermion model.
We compute
the Kohn stiffness constant describing the 
ground state conductivity at zero temperature
by exact diagonalization
on finite chains of various sizes. The main question that we
address here
is how the metal-insulator transition and 
the critical
phase of aperiodic non-interacting models are affected by the presence of 
many-fermion interaction. 

The significance of our studies can be also
viewed from a different perspective. Recently, there have been
many theoretical
investigations of the effects of fermion-fermion interaction on the problem of
persistent currents in mesoscopic rings\cite{DisInt1,DisInt2}.
These studies were
motivated by the fact that the free fermion theory underestimates the
magnitude of the observed persistent current compared to the observed
experimental value\cite{experiments}. The studies 
on spinless models\cite{DisInt1,DisXXZ}
showed that the repulsive interaction always decrease the amplitude of
the current. As pointed out by Giamarchi et al\cite{DisInt2},
both attractive and repulsive ground states have charge density fluctuations
which are easily pinned by the disorder. However, in the case of attractive
ground state, superconducting fluctuations screen the disorder
resulting in the enhancement of current. Our numerical results on
aperiodic interacting systems are consistent with the above picture.
Novel aspect of our results is the existence of a peak in the stiffness
constant at a characteristic value of the attractive interaction suggesting
the existence of a new type of phase in strongly correlated disordered
fermion systems.

In section II,  we describe the basic model under investigation
and give a short description of the
method utilized  to compute the charge stiffness $D_c$.  Section
III describes the results obtained from our numerical simulations.  Finally,
section IV contains our conclusions and a discussion of the
possible implications of this research.

\section{Model System and Charge Stiffness Calculation}
We studied an interacting spinless fermion model on a 1D ring in an
aperiodic potential,
\begin{equation}
H = - \sum_{i=1}^{N} (c_i^\dagger c_{i+1} + c_{i+1}^\dagger c_i) +
V \sum_{i=1}^{N} n_i n_{i+1} + \sum_{i=1}^{N} h_i n_i.
\label{spinless}
\end{equation}
The site dependent potential is chosen to be of the form,  
$h_i=\lambda \cos(2 \pi \sigma i^{\nu})$. Here,
$\lambda$ represents the strength of the potential and $\sigma$
is an irrational number which is chosen for convenience to be
the {\it Golden Mean} ($\frac{\sqrt{5}-1}{2}$).
The parameter $\nu$, determines the nature of aperiodicity:
for $\nu=1$  the potential is quasiperiodic
while for $\nu>3$  it has been shown to generate pseudorandom terms. Therefore,
this particular form of potential facilitates the
study of both quasiperiodic as well as pseudorandom cases
by varying $\nu$.

For $\lambda=0$, the interacting spinless fermion problem could
be mapped to the Heisenberg-Ising XXZ spin 
problem\cite{Jordan-Wigner}. This is an
old problem that have been extensively studied and for which
a closed Bethe's anzast solution exists\cite{XXZ}. 
In the non-interacting limit ($V=0$), the quasiperiodic case ($\nu=1$ )
can be reduced to 
the famous Harper equation\cite{Harper55}. The Harper equation
exhibits a metal-insulator transition
in one dimension\cite{Sokoloff,Hiramoto92}. 
At the
onset of transition $\lambda_c =1$, the quantum states are neither
extended nor
localized but instead exhibit fractal characters and have been termed
as {\it critical}.  The spectrum contains an
infinite number of gaps and is believed to be
a Cantor set of zero measure. These interesting aspects of the wave function
and the spectra 
have been shown to be reflected in the transport properties such as
Launder resistence\cite{Liu86,Cruz93}.

For finite values of fermion interaction $V$, 
the problem is more complicated due to the many-body
nature of the wave function. Unlike the non-interacting case, where
the behavior of the system can be described by studying 
the single particle wave function and the associated eigenvalues,
in many-body problem, one needs an alternative method to characterize the
nature of many-body state .
Recently, the Kohn stiffness
constant $D_c$ has been introduced to 
characterize the difference between the
metallic and insulating phases as it
gives a direct quantitative
measure of the electronic conductivity of 
the system\cite{Cstiffness}. In this paper, we will use $D_c$
to determine the nature of the phase of the aperiodic system.
 
To compute the Kohn stiffness constant $D_c$ , we assumed 
periodic boundary conditions for the fermion model described by 
equation~(\ref{spinless}). We are interested in the {\it persistent}
current response to a vector potential of magnitude $|\vec{A}| = 
\frac{\Phi}{N}$
in the $x$ direction, where $\Phi$ is the  flux threading the 
1D ring and N is the number of sites in the chain. We used the Lanczos
diagonalization method\cite{Diagonalization}
to obtain the ground state energy
of the system $E_0(\Phi)$ as a function of the flux. The Kohn
stiffness constant is then given by 
the equation\cite{DisInt1,Cstiffness},
\begin{equation}
D_c = \frac{N}{2} \frac{d^2 E_0(\Phi)}{d \Phi^2} |_{\Phi = \Phi_{min}}. 
\end{equation}
The numerical calculations were done using several sizes for
the 1D ring up to maximum size of $N=14$.  We studied the behavior
of the Kohn stiffness constant $D_c$  as a function of the parameters
$V$ and $\lambda$ describing the strength of the interaction
and the strength of the aperiodic potential respectively.

\section{Simulation results}
We did simulations of the aperiodic spinless system
For various sizes $N$, and the
electronic densities $\rho = \frac{N_e}{N}$, for many values of the
parameters $V$ and $\lambda$.
To simulate golden mean quasiperiodicity into the model,
we used Lanczos methods
for systems of various Fibonacci sizes. Furthermore, in order to keep
the fermion density $\rho$ 
$almost$ a constant, we worked with
densities  which are 
the rational approximants to the 
golden mean $\sigma$ or the square of the golden mean $\sigma^2$.
This procedure provides several possible sizes (5, 8, 13)
for which the Lanczos
diagonalization can be done at almost constant density.
Therefore, our studies are for systems away from half-filling
where the umklapp processes become irrelevant and the system in absence
of disorder is metallic.
By studying few different sizes, we were able to monitor
finite
size effects that could be present.

Figure 1, shows the results for $D_c$ describing the interplay between
aperiodicity (with $\nu=1$) and fermion-fermion interaction. 
The ground state transport properties are obtained for
both the repulsive as well as for the attractive many-body interaction.
The values of $D_c$
decrease as $\lambda$ increases for all values of the fermion interaction.
Consistent with the previous results on disordered systems\cite{DisInt1},
the repulsive interaction is found to decreases the $D_c$ while the attractive
interaction is found to increase the stiffness.
As seen from the figure, an interesting aspect of the attractive ground state
is the existence of a very prominent peak
around $V=-2.5$ that survives even in the regime where the non-interacting
system is an insulator. The location of the peak is insensitive
to the value of $\lambda$, however the peak gets narrower with
increase of $\lambda$. Similar effects were also observed for other
densities which are rational approximants of $\sigma^2$.

The figures indicate the possible existence
of a region where $D_c$ may takes intermediate values:
between those of a metallic 
and those corresponding to the 
Anderson localized insulating phase. In
order to study this effect in more detail, we performed further
simulations to study
the behavior of $D_c$ versus $\lambda$ for different values
of $V$ (see figure 1c).
The figure clearly shows that in spite of the general decreasing behavior
of $D_c$ with $\lambda$ for any value of $V$, for $V=-2.5$ the
values of $D_c$ are relatively larger than those for
$V=-1$ and $V=0$. This confirms the
special behavior of $D_c$ around $V=-2.5$.

Figure 2 shows the corresponding results for
the pseudorandom case ($\nu >3$).
For $\lambda=0$ case we observe the
well known transition  from an insulating to the metallic phase
as the
interaction strength parameter $V$ is varied.
In analogy with the results of the quasiperiodic case,
the attractive ground state in the pseudorandom case also
exhibits the novel characteristics
namely the existence of a peak in $D_c$ as $V$ is varied. However, unlike
the quasiperiodic case, the effect is less dominant in 
the pseudorandom. For the repulsive ground state, the interaction decreases the $D_c$
value. For larger values of $\lambda$ a slight increase is observed
for small values of $V$.

The intriguing behavior of the conductivity manifested by a characteristic peak
as the attractive interaction $V$ is varied, for various strengths of
the deterministic disorder $\lambda$ was further analyzed by studying
the dependence of $D_c$ on the size of the system. We notice that
even though the height of the peak in $D_c$ decreases with $N$, the
size of the system, this decrease is rather slow, particularly in comparison
with the variation in $D_c$ with $N$ in the regime far from the peak, i.e.
in the insulating phase. We conjecture that in the insulating phase,
the $D_c$ decays exponentially with the size of the system, while in
the regime near the peak, the charge stiffness decays as a power
law. This conjecture was verified at a special point $V=0$. For
$\lambda=2$
where the system is known to be in a critical state, the $D_c$ value
was found to exhibit a power law decay with the size of the system.
On the other hand, in the localized phase, the $D_c$ value was
shown to decay exponentially with $N$.

We would like to mention that the reasoning for the above conjecture
regarding the variation in $D_c$ with $N$ is in our observation that
figure 1 describing the dependence of $D_c$ with $V$ is reminiscent
of the plots of total band width (TBW) in the models of non-interacting
fermions in a quasiperiodic field.
It is known, for example,
that the TBW for a critical non-interacting
fermion model decays like a power law with the size of the
systems.  This is in clear contrast to the behavior at the
localized phase where the TBW decays exponentially.  This
fact offers a powerful criterion to distinguish both
phases.
Since the TBW
is a measure of transport properties, it is conceivable that
its scaling
properties are similar to those of the $D_c$ constant.

\section{Conclusions and Discussion}

The purpose of this work has been to gain some
insight in understanding of  
the transport properties of a $1D$ chain of spinless fermions
under the concurrent presence of strong interaction
and aperiodicity. The numerical Lanczos diagonalization method
used for these simulations is in essence 
an exact method\cite{Diagonalization}. 
Unfortunately,
its major limitation is due to practical
aspects; even for
relatively small sizes of the systems, large
amounts of RAM memory and fast CPUs are required to
obtain accurate values of the ground state energy.

In spite of these difficulties,
we think that is possible to extract some valuable information 
from our numerical simulations as our results are independent of various
parameters such as the amount of disorder, fermion density and the system
sizes.
First of all, our calculations confirm the known fact that
the effect of disorder 
in the interacting system is the localization of
the metallic
phase. However, the simulations
show that for aperiodic potentials, we may have 
a phase that is not completely
localized as in an Anderson insulator, nor completely metallic
as in a Mott conductor. This fact seems to be 
independent of the density. 
Comparison of quasiperiodic and pseudorandom cases suggest that
in the quasiperiodic case this intermediate phase may exist in a finite parameter interval
while in the pseudorandom case
this phase may exist at a single point marking the boundary between
the metallic and Anderson insulating phase. This distinction between
the behavior in the quasiperiodic and the pseudorandom cases is reminiscent
of the analogous distinctions that is known to exist in some
non-interacting models studied previously\cite{Satija94}.

Our preliminary results are on systems of rather small sizes,
and therefore, it is hard to reach any definite conclusion regarding the
the nature of the new phase proposed here. On the other hand, the existence of a region
with a characteristic peak is seen on systems of various sizes as well
as of various electron densities. The clarity with which this peak appears
seems to hint some new mechanism involving some sort of
competition other than the known screening of the disorder
due to superconducting fluctuations. Therefore,
it is tempting to speculate the possibility of a new type of
phase in aperiodic strongly correlated systems.
We hope that our preliminary studies will simulate further
research in this area.

\acknowledgements

The research of IIS is supported by a grant from National Science
Foundation DMR~093296.  JCC would like to thank the people at CIF 
(International Physics Center) in Bogot\'{a}
for their constant help and encouragement during his career
and acknowledge the support obtained through a
COLCIENCIAS-BID-ICETEX scholarship-loan from Colombia.

\begin{figure}
\caption{Quasiperiodic case ($\nu=1$). 
(a) Charge stiffness
versus $V$ for $\rho = \frac{3}{8}$
and $\sigma = \frac{5}{8}$.
(b) Charge stiffness
versus $V$ for $\rho = \frac{5}{13}$
and $\sigma = \frac{8}{13}$.
Different curves
represent different values of $\lambda$, namely:
$\lambda=0.0$  (solid line),
$\lambda=0.5$  (dotted line),
$\lambda=1.0$  (short-dash line),
$\lambda=1.5$  (long-dash line) and
$\lambda=2.0$  (dot-dash line).
(c) Charge stiffness versus $\lambda$
for $\rho = \frac{5}{13}$ and $\sigma = \frac{8}{13}$.
Different curves
represent different values of $V$, namely:
$V=-2.5$  (solid line),
$V=-1.0$  (dotted line) and
$V=0.0$  (short-dash line).}
\label{fig1}
\end{figure}

\begin{figure}
\caption{Pseudorandom case ($\nu>3$).
(a) Charge stiffness
versus $V$ for $\rho = \frac{3}{8}$.
(b) Charge stiffness
versus $V$ for $\rho = \frac{5}{13}$.
Different curves
represent different values of $\lambda$, namely:
$\lambda=0.0$  (solid line),
$\lambda=0.5$  (dotted line),
$\lambda=1.0$  (short-dash line),
$\lambda=1.5$  (long-dash line) and
$\lambda=2.0$  (dot-dash line).}
\label{fig2}
\end{figure}


\begin{references}

\bibitem{email1}e-mail: jchaves@gmu.edu.

\bibitem{email2}e-mail: isatija@sitar.gmu.edu.

\bibitem{Anderson79} E. Abrahams, P. W. Anderson,
D. C. Licciardello and T. W. Ramakrishnan,
Phys. Rev. Lett. 42, 673 (1979).

\bibitem{Sokoloff} J. B. Sokoloff, Phys. Rep. 126, 189 (1985).

\bibitem{Hiramoto92} H. Hiramoto and M. Kohmoto, Int. J. Mod. Phys. B 6, 281
(1992).

\bibitem{Satija94} I. I. Satija, Phys. Rev. B 48, 3511 (1993);
Phys. Rev. B
49, 3391 (1994).

\bibitem{Aperiodic} S. Das Sarma, S. He and X. C. Xie, Phys. Rev.
B 41, 
5544 (1990); J. C. Chaves, I. I. Satija
and J. A. Ketoja,  Physica Scripta 52, 614 (1995).

\bibitem{Liu86} Y. Liu and K. A. Chao, Phys. Rev. B 34, 5247 (1986).

\bibitem{Cruz93} H. Cruz and S. Das Sarma, J. Phys. I France 3,
1515 (1993).

\bibitem{DisInt1} G. Bouzerar, D. Poilblanc and G. Montambaux,
Phys. Rev. B 49, 8258 (1994);
M. Abraham and R. Berkovits, Phys. Rev. Lett. 70, 1509 (1993). 

\bibitem{DisInt2} T. Giamarchi and B. S. Shastry, Phys. Rev. B 51,
10915 (1995).

\bibitem{experiments}  L.P. Levy, G. Dolan, J. Dunsmuir and
H. Bouchiat, Phys. Rev. Lett. 64, 2074 (1990); V. Chandrasekhar,
R.A. Webb, M.J. Brady, M.B. Ketchen W.J. Galager, 
and A. Kleinsasser, Phys. Rev. Lett. 67, 3578 (1991); D. Mailly, 
C. Chapelier, and A. Benoit, 
Phys. Rev. Lett. 70, 2020 (1993); D. Mailly, 
C. Chapelier, and A. Benoit, Physica B, 197, 514 (1994).

\bibitem{DisXXZ} W. Apel, J. Phys. C 15, 1973 (1982);
T. Giamarchi and H. J. Schulz, Phys. Rev. B 37, 
325 (1988); W. Lehr, Z. Phys. B 72, 65 (1988);
R. Shankar, Int. Jour. Mod. Phys. B 4, 2371 (1990);
H. Pang, S. Liang and J. F. Annett, Phys. Rev. Lett. 71, 4377 (1993).

\bibitem{Jordan-Wigner} E. Fradkin, 
{\it Field theories of condensed matter systems},
Addison-Wesley Pub. Co., Redwood City, Calif. 1991;
A. Tsvelik, {\it Quantum field theory in condensed matter physics},
Cambridge University Press,  New York, NY, USA  1995.

\bibitem{XXZ} B. Sriram Shastry and B. Sutherland, 
Phys. Rev. Lett. 65, 243 (1990); J. D. Johnson,
J. Appl. Phys. 52, 1991 (1981).

\bibitem{Harper55} D. R. Hofstadter, 
Phys. Rev. B 14, 2239 (1976); P. G. Harper,
Proc. Phys. Soc. London  A 68,
874 (1955).

\bibitem{Cstiffness} D. J. Scalapino, S. R. White and S. Zhang,
Phys. Rev. B 47, 7995 (1993); Q. P. Li and X. C. Xie, 
Phys. Rev. B 49, 8273 (1994).

\bibitem{Diagonalization}  H. Q. Lin and J. E. Gubernatis,
Comput. Phys. 7, 400 (1993); H. Q. Lin,
Phys. Rev. B 42, 6561 (1990); H. H. Roomany,  H. W. Wyld
and L. E. Holloway, Phys. Rev. D 21, 1557 (1980).

\end{references}
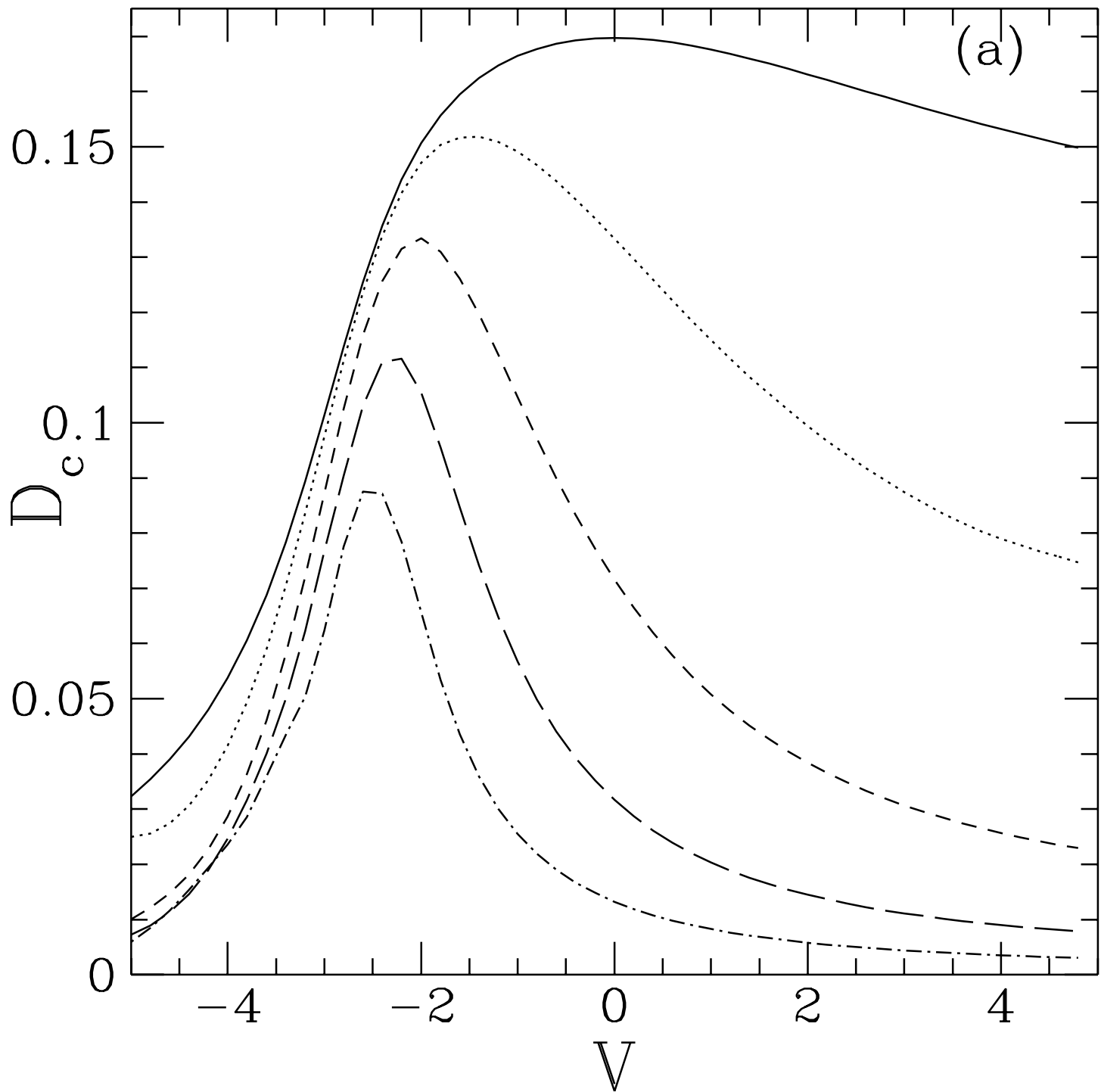
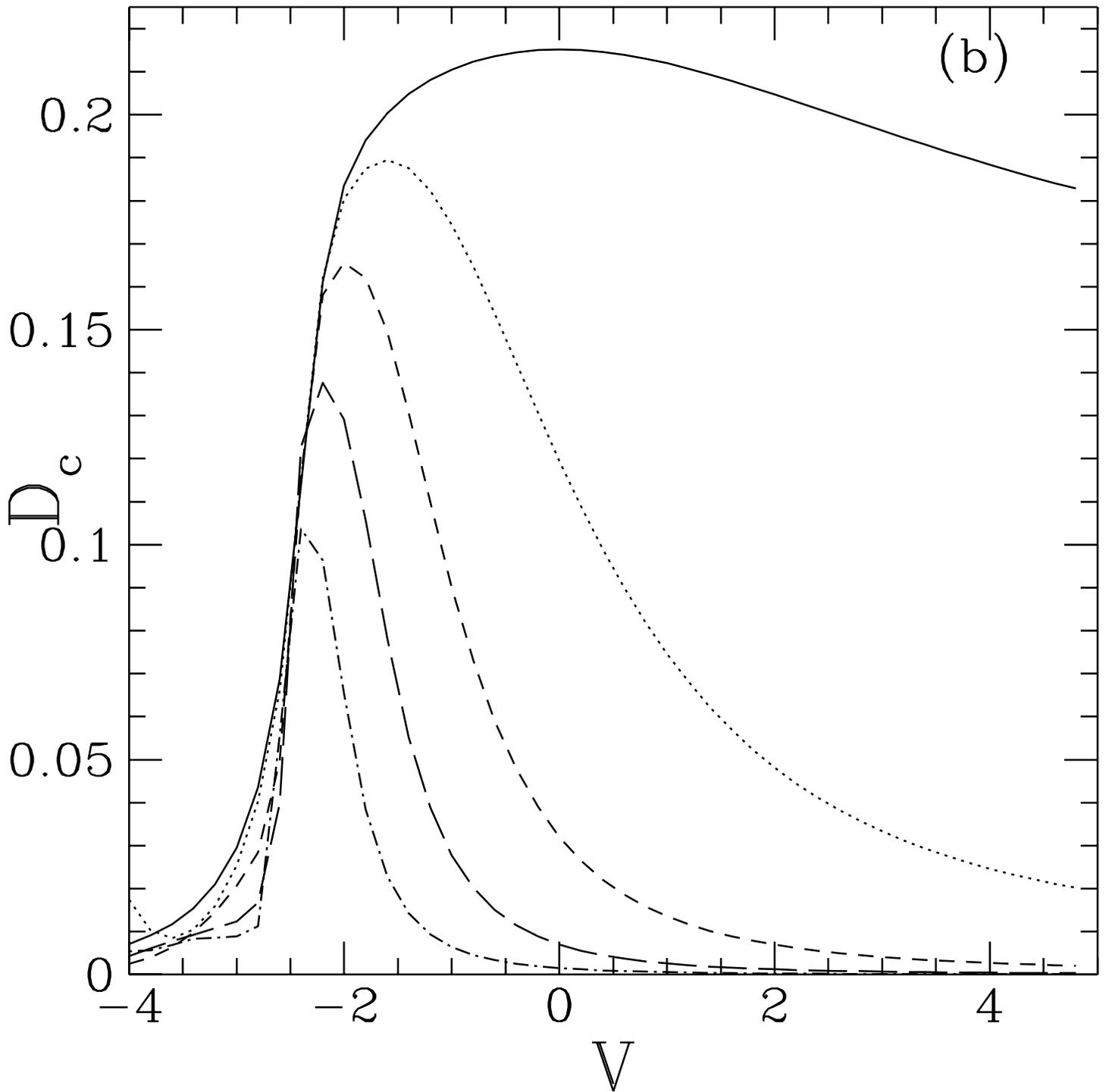
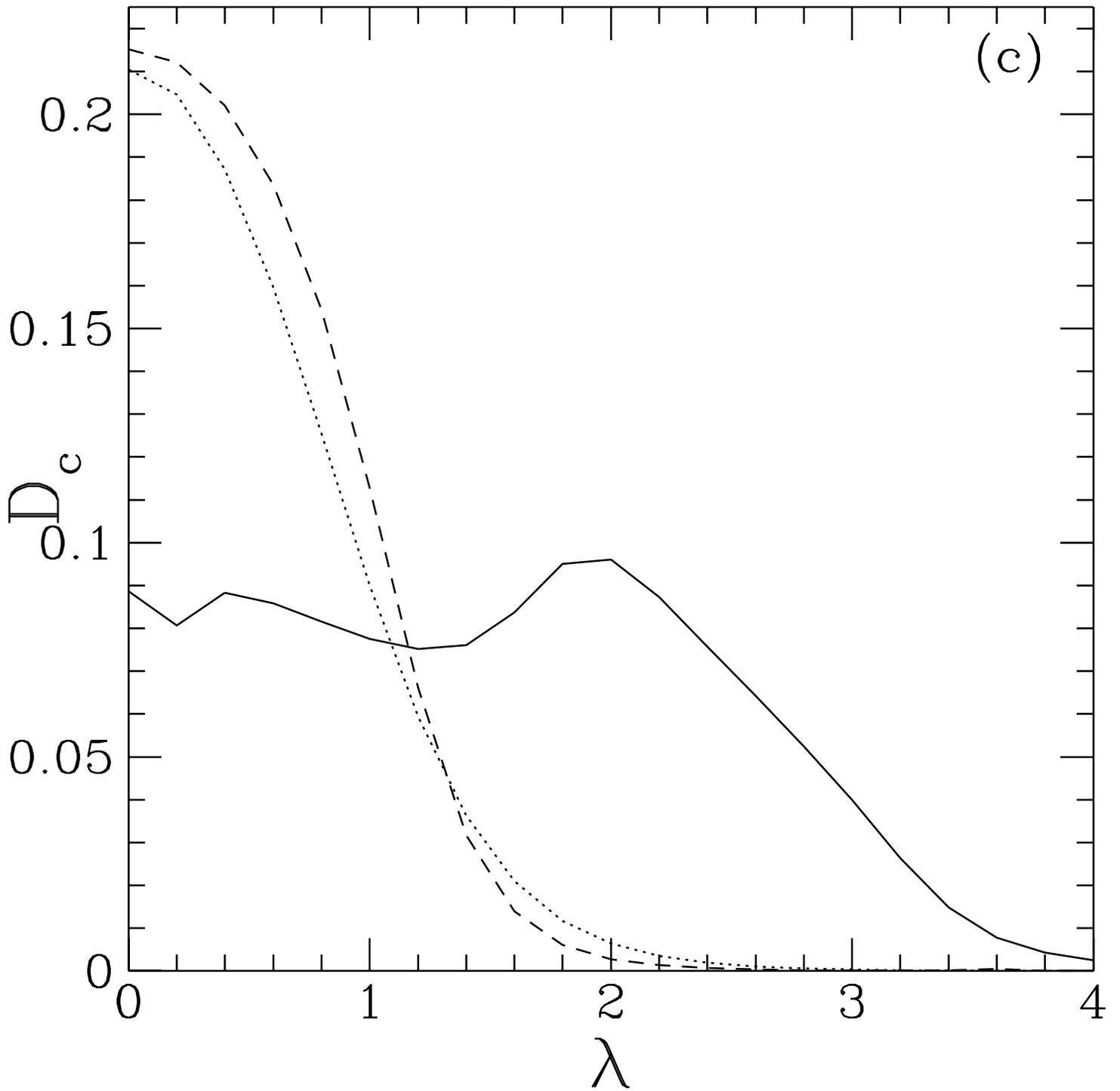
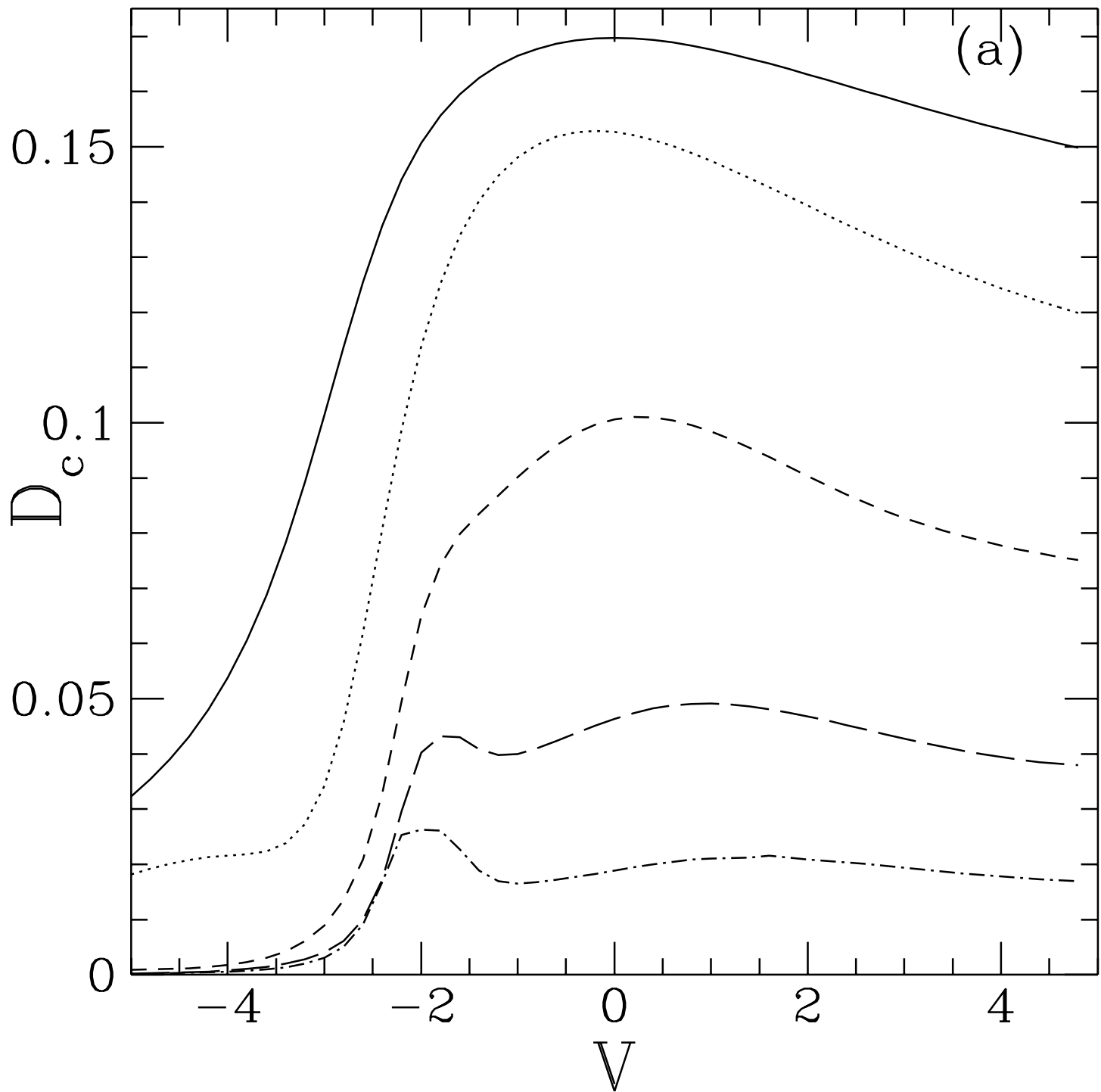
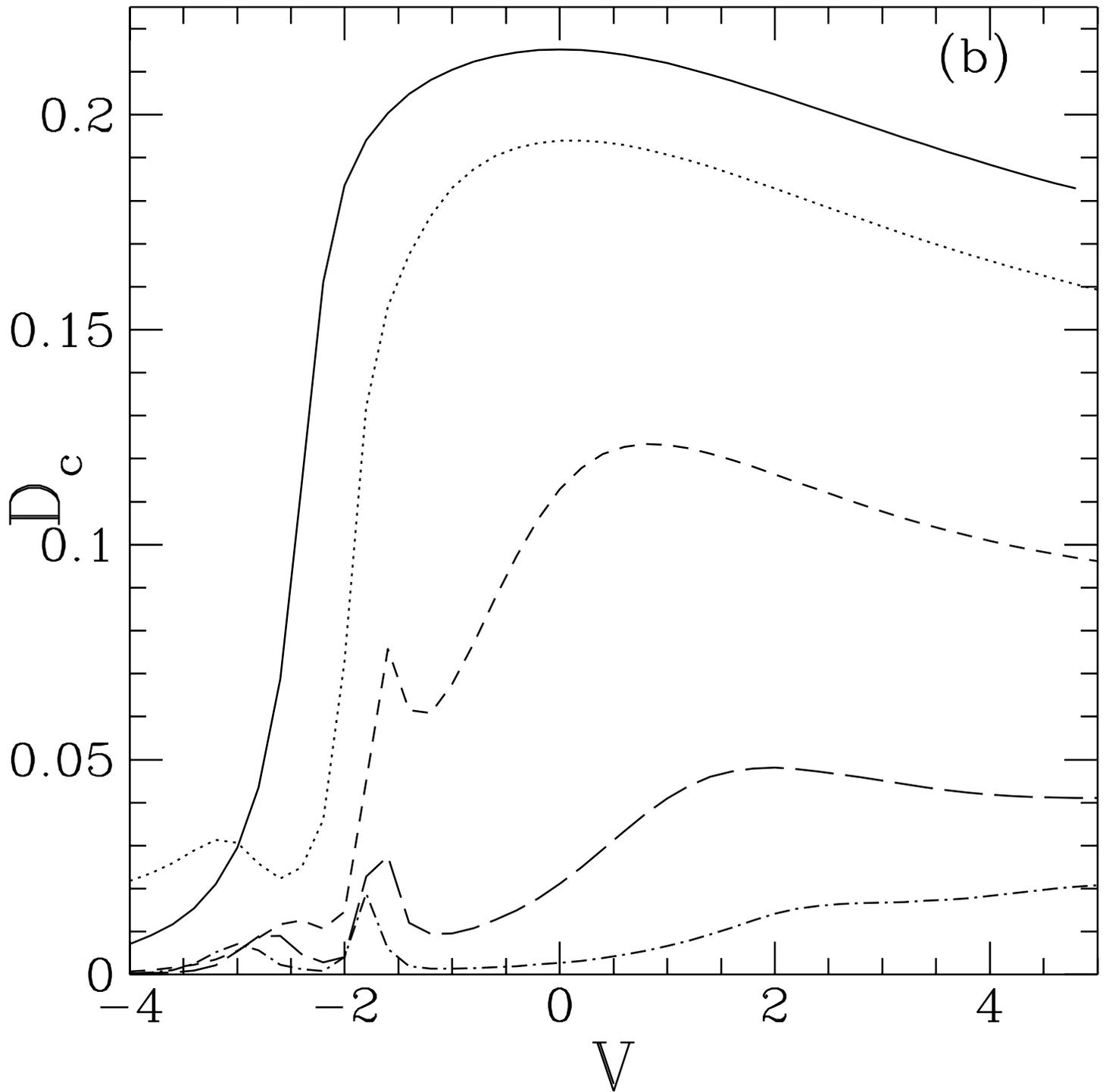
\end{document}